# Architecture of A Scalable Dynamic Parallel WebCrawler with High Speed Downloadable Capability for a Web Search Engine

**Debajyoti Mukhopadhyay**[1,2]  **Sajal Mukherjee**[1]  **Soumya Ghosh**[1]  **Saheli Kar**[1]  **Young-Chon Kim**[2]

[1] Web Intelligence & Distributed Computing Research Lab, Techno India (West Bengal University of Technology)
EM 4/1, Salt Lake Sector V, Calcutta 700091, India
Emails: {debajyoti.mukhopadhyay, sajal.mukherjee, soumya.ghos, saheli.kar}@gmail.com

[2] Chonbuk National University, Division of Electronics & Information Engineering
561-756 Jeonju, Republic of Korea; Email: yckim@chonbuk.ac.kr

**ABSTRACT**
*Today World Wide Web (WWW) has become a huge ocean of information and it is growing in size everyday. Downloading even a fraction of this mammoth data is like sailing through a huge ocean and it is a challenging task indeed. In order to download a large portion of data from WWW, it has become absolutely essential to make the crawling process parallel. In this paper we offer the architecture of a dynamic parallel Web crawler, christened as "WEB-SAILOR," which presents a scalable approach based on Client-Server model to speed up the download process on behalf of a Web Search Engine in a distributed Domain-set specific environment. WEB-SAILOR removes the possibility of overlapping of downloaded documents by multiple crawlers without even incurring the cost of communication overhead among several parallel "client" crawling processes.*

## 1. INTRODUCTION

A crawler also popularly known as spider or robot is a program which visits Web servers spread across the world irrespective of their geographical locations, downloads and stores Web documents on a local machine mostly on behalf of a Web Search Engine.[1][2][3] The functionalities of a Web crawler is given below:

- The crawler starts crawling with a set of URLs fed into it, known as seed URLs.
- The crawler downloads the page.
- It extracts the URLs from the downloaded page and inserts them into a queue. From the queue the crawler again retrieves the URL for downloading next pages.
- The downloaded page is saved in the repository.
- The process continues until the crawler stops.

In order to maximize the download rate, if multiple crawling processes are employed in parallel then the resulting crawler is termed as a "parallel crawler."[1][6]

While designing a dynamic parallel crawler we faced the following design challenges:

- Preventing the overlapping of Web-pages among concurrently running crawling processes i.e., no two crawlers should download and store the same Web document.
- Maintaining the high quality of downloaded documents during the crawling process.
- Minimizing the communication overhead among the concurrently running crawling processes, that is evident in any parallel crawler implementation.

Here we discuss in brief the basic idea of parallel crawlers. Then we will look into architectures of static crawlers, their drawbacks and overcoming those drawbacks with the help of our dynamic parallel crawler.

## 2. GENERAL CONCEPT OF PARALLEL CRAWLERS

A parallel crawler is implemented by means of more than one crawling processes, each individually downloading and storing pages. In order to co-ordinate the activities of these crawling processes, there must be some form of communication between them. Depending upon the method of co-ordination, parallel crawlers can be classified into two categories: (1) Static and (2) Dynamic.

### 2.1 Static crawlers

In case of these crawlers the Web is partitioned and assigned statically to individual crawling processes and each one of them is aware of others' assignments i.e., which crawling process is responsible for a particular Web document [1,6,7]. In this class of parallel crawlers, there is no central controller module to co-ordinate the activities of individual crawling processes.

*Independent crawler:* Crawlers may crawl independently without communicating with each other. All the decisions of crawling have to take by the individual crawler. After downloading and parsing a page the crawler may find an URL that belongs to another crawler. The crawler may handle this in two ways:

    It may discard the URL and continue to crawl on its own partition erasing any possibility of overlap. This mode is termed as the Firewall mode.[1] Here many important URLs will be lost reducing the quality of the ultimate search result.

    Second way to overcome this problem is that the crawler will download the page of the other partition which is termed as the Cross-over mode. It may result in downloading a single page more than once i.e., overlapping of a page will occur.

*Communicating crawler:* In the Exchange mode, there is an intercommunication link between every two crawler. When a crawler gets an URL that belongs to other partition it sends it to the crawler which is responsible for downloading that page. There are certain drawbacks of this design.

    Communication overhead is the first drawback of communicating crawler. As the number of crawlers increases the need of communication links will also increases, causing a communication overhead. Say, for example, a crawler c1 after visiting a page gets URLs of crawler c2, c3 & c4 respectively. First it will send the URL of c2 to c2 and then similarly to c3 and c4. In the meantime the crawler c1 pauses its crawling until the communication is complete. Thus we see that there is also a delay in the process.

### 2.2 Dynamic crawlers

In case of "dynamic crawlers" a central controller module divides the Web into a number of logical partitions and dynamically assigns each partition to an individual crawling process.[1][6] Here the controller provides each crawler with a set of seed URLs from which it starts crawling.

## 3. OUR APPROACH

Due to the enormous size of the Web we would naturally want to visit high quality popular pages first. Initially we have considered the back-link count as the quality metric i.e., we would like to visit pages in the order of their back-link counts. [2][3][8]

### 3.1 Partitioning Scheme:

We have partitioned the Web on the basis of the domain extensions such as *.com, .edu, .net, .biz* etc. A crawler can also visit a set (which we are calling DSet) of such domains. Say for example A crawler can crawl a domain-set or DSet D:{*.net, .biz*}. Note that each of our crawling processes which we named as "Crawl-clients" crawl a single DSet in its lifetime i.e., there is no exchange of partitions. The functional architecture of WEB-SAILOR is shown in Figure1. Each Crawl-client receives seed URL for its DSet, loads the Web-page, parses the page for outbound links and stores them for future consideration and finally saves the downloaded page into local repository for indexing. As the whole Web is partitioned among the crawlers domain-wise, there is no question of overlapping. Sometimes a Web page of a particular domain may be populated with links to several Web pages of some other domain. Most practical example could be a Website of online bookstore in *.com* domain (e.g., www.amazon.com) which is very popular in *.edu* domain as many universities publish booklists with links to such online book store. In such cases, using static crawlers will either degrade the ultimate search result quality (in case of Independent Crawlers, as a popular page is discarded) or crawl decision quality will be deteriorated(in case of Communicating Crawlers, as visiting a popular page is delayed). This is where the concept of Seed-server comes in.

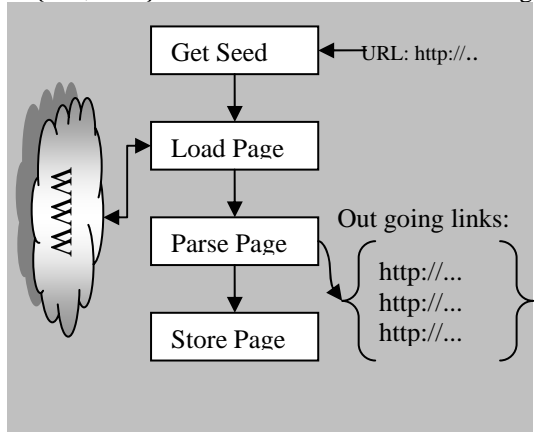

**Figure 1:** Functional view of Crawl-client

due to the fact that in such systems no single crawler has the complete view of the entire Web. In our design, higher level Seed-server(s) have the global image of the Web and they will make the crawling decision instead of the clients. In our proposed architecture of WEB-SAILOR, depicted in Figure2, each Crawl-client crawls different partitions i.e., different DSets. It receives seed URLs from Seed-server and then sends the parsed outbound URLs back to the server without following them directly. The Seed-server maintains a DSet-wise global index of back-link counts for URLs which we termed as the "URL-Registry" and updates it as more outbound URL references come from Crawl-Clients. If available (i.e., the page has been crawled previously), server can also use other ranking indexes [8][9][10].

### 3.3 URL-Registry Structure
In our design, we have built a DSet-wise, global, central data-structure i.e., the URL-Registry which is maintained at the server to keep track of

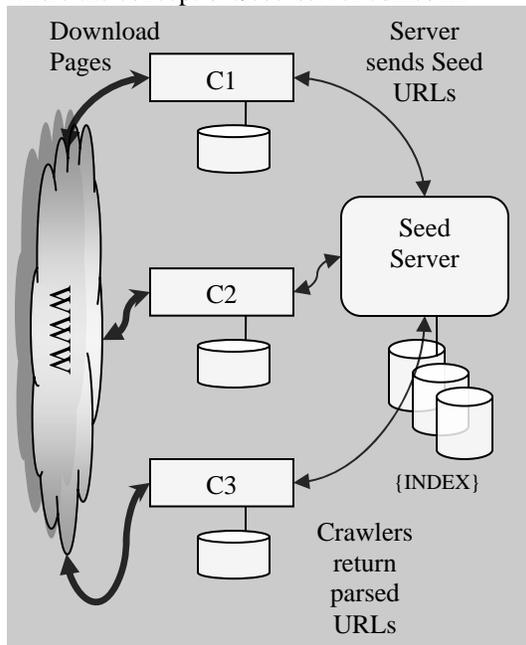

**Figure 2:** Crawling Architecture of WEB-SAILOR

### 3.2 Server Centric Approach
Static parallel crawlers suffer from poor crawling decision i.e., which Web-pages to crawl next,

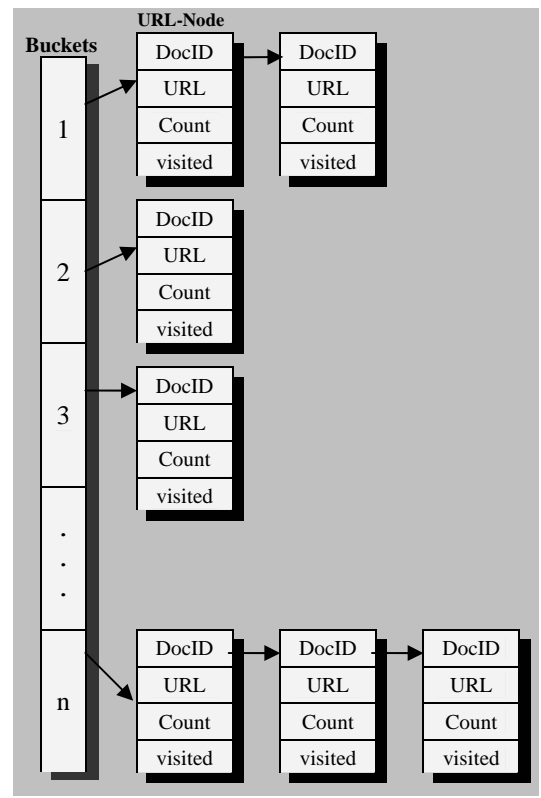

**Figure 3**: URL-Registry Structure

all the URLs of that particular DSet and their back-link counts. The URL-Registry is a hash-based structure where a hash function is used to generate a unique DocID of each URL.[3] The URL-Registry consists of *n* number of buckets each bucket consists of 0 or more number of URL-Nodes as given in the Figure3. Modular division of DocID by *n* obtains the appropriate bucket for a particular URL-Node. Each URL-Node consists of 4 fields:

- *DocID*: Stores the unique value generated by using the hash function on the URL.
- *URL*: Stores the URL as a string.
- *count*: Stores the back-link count of the URL. The count is incremented each time it is referred by a Web page.
- *visited:* This field is marked if the page corresponding to the URL is visited.

We have maintained a separate URL-Registry for each DSet. This increases the parallelism as multiple URLs can be retrieved from different URL-Registry structures corresponding to different DSets Another advantage of the URL-Registry structure is that we can decrease the time required for searching a particular node by increasing the value of *n*, i.e., the number of buckets. As the structure expands vertically, the horizontal size for the buckets decreases resulting in a decrement in linear search, which is time consuming.

## 4. ADVANTAGES OF OUR APPROACH

This server centric approach has several advantages such as improved crawling decision, less network load on Web servers, dynamic load balancing etc. We will explain them below.

### 4.1 Improving Crawling Decision:

In this system, the Seed-server has the quality metric i.e., the back-link counts for all the pages in all the partitions. Hence the crawling decision quality will always be as good as that of a single crawler but the pages will be downloaded concurrently.

### 4.2 Minimizing Web Server Load

A large scale crawler comes with a responsibility of not overloading any Web server. It has been observed that a Web-page contains a significant amount of links pointing towards its own Web server. Crawl-clients send all outbound links from a page to the Seed-server and Seed-server sends most popular unvisited links to clients as seed. This popularity is measured globally among all known URLs of all partitions. We consider it improbable that a large number of pages of a Web server to have same popularity. Hence it is unlikely that a number of pages of same Web server will be downloaded concurrently.

### 4.3 Dynamic Load Balancing

In our design the Seed-Server can globally monitor performance of each Crawl-client of all partitions. We can take the number of seed URLs pointing towards a partition as its popularity metric. Each client maintains a number of connections to download pages concurrently. If server notices that it has less (than a threshold number of) seed for a particular DSet it will send a message to the Crawl-client corresponding to that DSet to *"slow down"* and the client will reduce the number of parallel connections. Similarly when the server sees a huge amount of seeds for a particular DSet it asks the corresponding Crawl-client to *"hurry up"* i.e., to increase the number of concurrent connections. These phenomena are shown in Figure4a and Figure4b respectively.

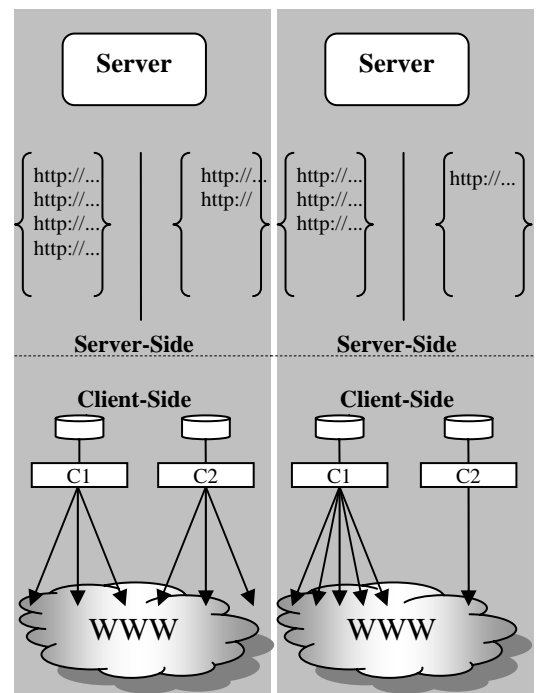

**Figure 4a:** Connections before Load Balancing    **Figure 4b:** Connections after Load Balancing

### 4.4 Scalability

Our approach is inherently more scalable than other communicating crawlers as in case of this genre of crawlers, a crawling process must communicate with all other crawling processes as necessary.[4] If there are $N$ numbers of such processes running, they will require total $N!$ connections to communicate to each other. Addition of more crawlers at run time is also difficult as the new one must notify all other existing crawling processes. In our design, $N$ number of crawling processes needs to maintain only $N$ connections with the Seed-server and the addition of a new Crawl-client is only visible to the seed-server. Of course there will be a limitation on the number of Crawl-clients under a single server depending on its hardware capability. In order to scale-up the system further one may use the same structure recursively. The design is shown in Figure5 with two additional servers. Seed server $S_1$ & $S_2$ is serving $N$ & $M$ clients respectively, where $C_{ij}$ implies $j^{th}$ Crawl-client of the $i^{th}$ Seed-server. If $S_2$ receives an URL which does not fall into the DSets of its own clients, it will send it to the higher level server $S_{12}$ which in turn route it to appropriate seed server $S_1$.

### 5. EXPERIMENTAL SETUP

In our prototype we have used two machines as Crawl-clients and one as Seed-server. One client was used as client for *.com* domain; the other was used for downloading pages of *.edu, .net, & .org*. We varied the number of initial connection for each Crawl-client to observe the performance of the server. Also, the Crawl-client crawling the *.com* domain was given more number of initial connections considering the huge number of pages in *.com* domain. We made the server multithreaded so that it can handle client requests (for dispatching seeds & receiving links) concurrently. Figure6 shows three different graphs, one for Client 1 with 25 connections, other for Client 2 with 10 connections and there is one showing total performance. We also added a third client in the system at runtime to see the effect on the server. Rate of number of pages downloaded remain steady in all the cases.

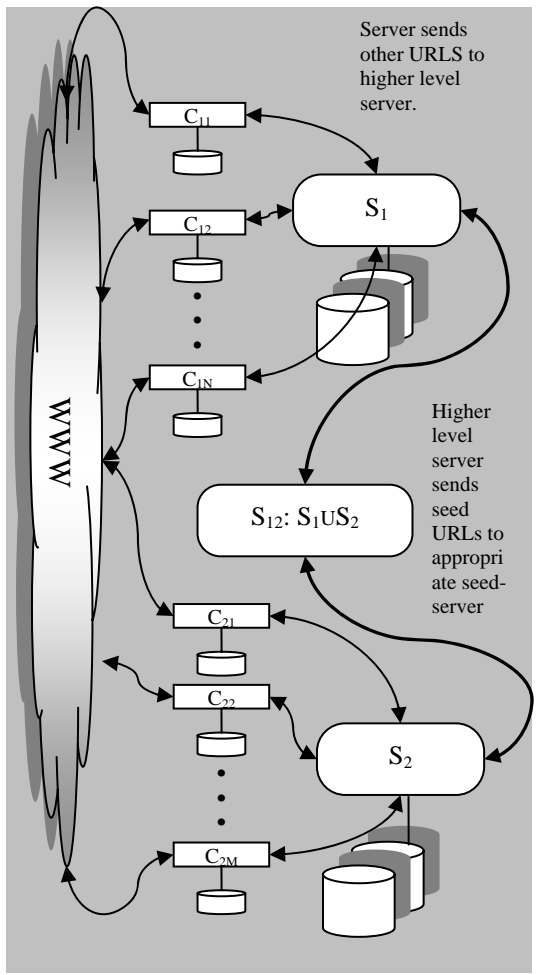

**Figure 5:** Scalable Design for very large scale crawling system

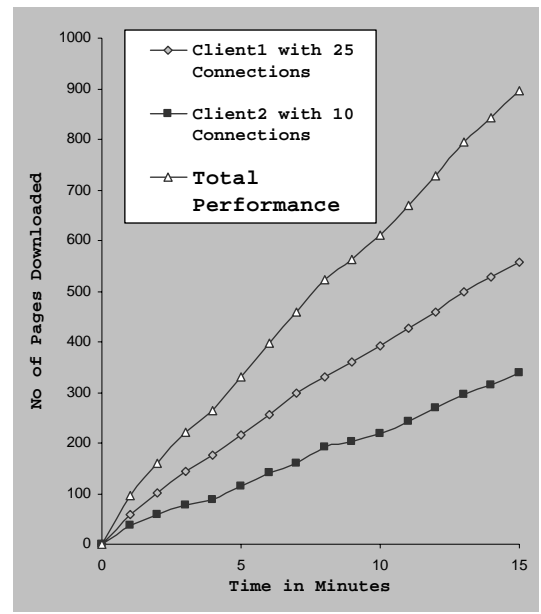

**Figure 6:** Performance of the Crawling System

## 6. CONCLUSION & FUTURE WORK

Parallel Crawler is an integral part of any large-scale Search Engine design. But due to of the secrecy maintained by the industry, little has been known about the modern trends of research in this area. Our paper is an effort to highlight a new Server-centric, dynamic, scalable approach of designing a parallel crawler which has been able to solve a number of problems that are inherent in any parallel crawler.

Firstly, our approach has been able to remove the possibility of overlapping of Web pages downloaded by different Crawl-clients. In our design, all the crawl decisions are made by the Seed-server only. The multithreaded Seed-server sends unvisited URLs as seed from a domain set-wise URL-Registry maintained by the Seed-server. So there is no question of redundant downloading by the Crawl-clients.

Secondly, we have been able to reduce the network load considerably resulting in better utilization of communication band-width. In our approach, since all the communication takes place between the Seed-server and the Crawl-clients only and the Crawl-clients don't need to communicate among themselves at all, so the number of communication paths is much less and adding new clients during run time incurs less overhead.

Thirdly, our architecture is not dependant upon any particular ranking scheme. Since the crawl decisions are solely made by the server, so it can accommodate any combination of ranking mechanisms and/or any other quality metric such as freshness of pages to maintain the high quality of downloaded pages. This feature adds to flexibility of our approach.

Lastly, this design offers a highly scalable approach towards designing parallel crawlers. Due to the self-recursive nature of our design, with the availability of adequate amount of resources, the prototype we have built can be extended hierarchically that will definitely result in a parallel crawler that can exploit a greater degree of parallelization. We are planning to carry on this as a part of our future work.